\documentclass[aip,reprint,twocolumn]{revtex4-2}

\usepackage{graphicx}
\usepackage{dcolumn}
\usepackage{bm}
\usepackage{color}
\usepackage[table]{xcolor}
\usepackage{braket}
\usepackage{amsmath}
\usepackage{amssymb}
\usepackage{graphicx}
\usepackage[unicode=true,pdfusetitle,
 bookmarks=true,bookmarksnumbered=false,bookmarksopen=false,
 breaklinks=false,pdfborder={0 0 0},pdfborderstyle={},backref=false,colorlinks=true]
 {hyperref}
\hypersetup{
 pdfborderstyle={},citecolor=blue}

\usepackage[utf8]{inputenc}
\usepackage[T1]{fontenc}
\usepackage{mathptmx}

\definecolor{BrickRed}{RGB}{132,31,39}

\begin{document}

\title{Disentangling Memory and Nonlinearity in Time-Multiplexed Optical Reservoir Computing}

\author{Elias R. Koch}
\email[Corresponding author: ]{elias.koch@uni-muenster.de}
\affiliation{Institute for Theoretical Physics, University of Münster, Wilhelm-Klemm-Str. 9, 48149 Münster, Germany }
\author{Julien Javaloyes}
\affiliation{Departament de Física and IAC$^{3}$, Universitat de les Illes Balears,
Campus UIB 07122 Mallorca, Spain }
\author{Svetlana V. Gurevich}
\affiliation{Institute for Theoretical Physics, University of Münster, Wilhelm-Klemm-Str. 9, 48149 Münster, Germany }
\affiliation{Center for Data Science and Complexity (CDSC), University of Münster,
Corrensstr. 2, 48149 Münster, Germany }
\author{Lina Jaurigue}
\affiliation{Institute of Physics, Technische Universitat Ilmenau, P.O.Box 100565, D-98684 Ilmenau,
Germany}

\begin{abstract}We analyze and disentangle the individual and combined roles of nonlinearity, transient dynamics, and delayed feedback, and investigate how these mechanisms contribute to different task requirements in photonic reservoir computing. This is achieved using a passive linear photonic reservoir in which a photodiode at the optical-to-electrical interface provides the sole source of nonlinearity. We demonstrate how the requirements of different tasks can be quantified and compared to the dynamics provided by the reservoir by explicitly calculating and comparing the contributing monomials. Our findings include that a reservoir with uncoupled virtual nodes and a quadratic nonlinearity can already achieve good performance for tasks derived from dynamical systems with quadratic nonlinearities, such as the Lorenz63 system. Furthermore, transient coupling and delayed feedback significantly enhance the computational capabilities by compensating for missing higher-order monomials through an effective multi-step integration scheme, given the minimum nonlinearity required for the task is present in the reservoir.
\end{abstract}

\maketitle

\section{Introduction}
\begin{figure}[t]
	\includegraphics[width=1\columnwidth]{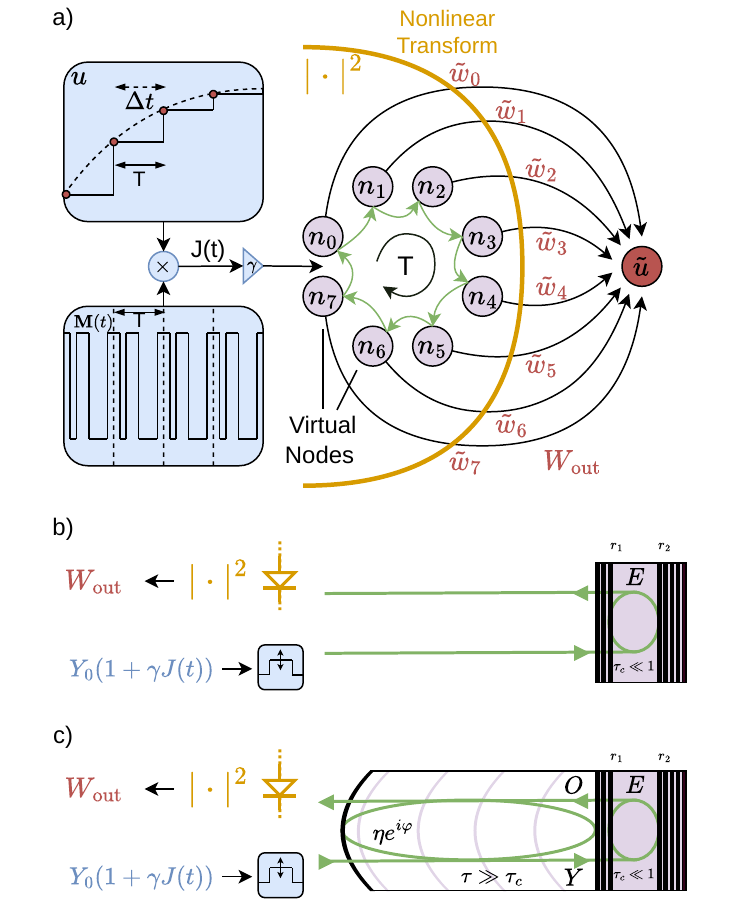}
	\caption{Comparing a general sketch of time-multiplexed single-node reservoir computing to our setup. Corresponding elements are colored accordingly. a) Sketch of a reservoir computer, using a single node as a reservoir through multiple virtual nodes introduced by the time-multiplexed input signal. b) Setup of a small optical micro-cavity (single node), injected with a monomode injection beam that can be modulated with the input data. The output is collected by a photodiode (PD), acting as a quadratic nonlinearity.
	c) The optical micro-cavity (single node) is now coupled to a long external cavity with a roundtrip $\tau$ acting as a feedback loop (coupling).}
	\label{fig:1}
\end{figure}


Over the last years \emph{reservoir computing} (RC) has emerged as a promising fast and efficient machine learning paradigm. Instead of optimizing the entire network for a specific task, RC employs a fixed nonlinear dynamical system, the reservoir, which projects an input into a high-dimensional state space, while only an output layer is trained, typically with linear regression \cite{J_GMD_01,LJ_CSR_09}. This drastically reduces the computationally expensive and slow training associated with classical recurrent neural networks (RNNs) \cite{TYH_NeuralN_19} while also mitigating overfitting. The reservoir will respond differently, depending on the type and order of input signals, just as each object thrown into a pond of water would induce different patterns on the surface~\cite{AMS_JJApp_24} maintaining an echo or trace of the input history in its current state. 

Typical candidates for RCs are networks of interacting nonlinear nodes or continuous media \cite{TYH_NeuralN_19}, while another type employs reservoirs consisting of a single physical node subjected to \emph{time-delayed feedback}~\cite{ASS_NC_11}. In the latter, $N$ so-called \emph{virtual nodes} can be created in the time-domain by multiplying each input point $u$ with a mask $\textbf{M}(t)$, resulting in $N$ linear combinations of that point. These linear combinations are subsequently injected into the reservoir as an input signal, each for a certain amount of time each, before an output value is recorded. This results in the $N$ virtual nodes (cf. $n_1,...,n_N$ in Fig.~\ref{fig:1}~(a)), which play an analogous role to the physical nodes in a network-type reservoir. The injection time per linear combination is the so-called \emph{node separation} $\theta$ and a new input point is applied after each \emph{input clock cycle} $T=N\cdot \theta$, see Fig.~\ref{fig:1}~(a). This concept has enabled numerous experimental implementations, particularly in photonics, owing to high operating speed and amenable hardware implementation~\cite{BSMF_NC_13,VMV_NC_14}.
Some realizations are useful candidates for applications in real-time information processing such as speech recognition \cite{PDS_SRep_12, LSB_OptEx_12, LBM_PRX_17, CYK_OE_19}, nonlinear channel equalization \cite{PDS_SRep_12, LSB_OptEx_12, CYK_OE_19,ACG_IEEE_20} or chaotic time series prediction \cite{LSB_OptEx_12, BSMF_NC_13, KNTU_OE_18}.

In particular, both active and passive photonic reservoirs have been considered. Here, active reservoirs are based on active components such as semiconductor lasers~\cite{BSMF_NC_13,KNTU_OE_18,CYK_OE_19,HS_Photonics_19} or networks of semiconductor optical amplifiers (SOAs). Their operation usually demands for additional energy supply for the reservoir on top of the input data signal. Although these systems have demonstrated excellent performance on a variety of benchmark tasks, they come with rather complex setups, careful parameter tuning and comparably high energy consumption~\cite{SNS_NanoP_17,ZV_IEEEA_23}.
As energy efficiency is one of the key advantages of RC, \emph{passive photonic reservoirs} have attracted increasing attention, including implementations based on passive components and resonators such as optical fiber cavities, microring resonators, and integrated photonic networks~\cite{VMV_NC_14,VDS_Optica_15,KFB_CoCo_17,PVM_FiP_19,NTH_CommP_21,DMM_OE_22,BHGB_OE_24}. Besides their substantially lower energy consumption, passive optical components typically offer a higher bandwidth, making them attractive for high-speed and parallel information processing \cite{SNS_NanoP_17}.

To fully harness these advantages, Vandoorne et al. implemented a purely \emph{linear} integrated network-type reservoir consisting of optical waveguides, splitters, and combiners~\cite{VMV_NC_14}. There, the necessary nonlinearity was provided by a \emph{photodiode} (PD) in the readout layer, performing a quadratic transformation of the optical field \cite{SNS_NanoP_17}. A conceptually similar single-node realization with a PD as only source of nonlinearity was realized by Vinckier et al., using a linear passive fiber cavity~\cite{VDS_Optica_15} as a reservoir. 
However, these setups cannot be operated all-optical~\cite{BSMF_NC_13} since the PD constitutes an optoelectronic interface~\cite{PDS_SRep_12,SNS_NanoP_17}.
Nevertheless, these studies showed that a purely linear optical reservoir combined with a nonlinear readout or input layer can successfully solve a wide range of benchmark tasks, while combining ultra-low energy consumption, conceptual simplicity and high performance \cite{VDS_Optica_15,SNS_NanoP_17}. 

In this paper, we explore another advantage of this type of architecture, namely its potential to disentangle and study the individual contributions of \emph{nonlinearity}, \emph{transient dynamics}, and \emph{time-delayed feedback}. In contrast to conventional nonlinear reservoirs, where transient dynamics and time-delay jointly enhance complexity and spread nonlinearity, the proposed linear architecture isolates these contributions. A deeper understanding of these mechanisms represents an important step towards identifying the conditions that determine RC performance and enable more informed choices in reservoir design. 

Our setup consists of a \emph{passive linear micro-cavity} enclosed by two distributed Bragg reflectors (DBRs), which can be coupled to an external \emph{delayed feedback loop} with a feedback mirror. Since the optical cavity itself remains linear, the only source of nonlinearity originates from a PD at the optical-to-electrical interface, though higher-order nonlinearities are also considered, cf. Fig. \ref{fig:1}~b). As these RC realizations are already optoelectronic, higher-order nonlinearities are not a purely theoretical construct, but could be realized e.g. with an additional electronic nonlinear transformation. As a consequence our setup allows to systematically control
\begin{itemize}
	\item the nonlinear transformation through the output layer;
	\item transient coupling through the virtual node separation $\theta$; and
	\item time-delayed feedback through the feedback strength $\eta$ and delay $\tau$.
\end{itemize}
Asides from providing insight into this specific implementation, such a decomposition addresses a broader question in reservoir computing: can we understand how and to what extent different effects can enhance computational performance? 

Although numerous performance measures have been proposed, recent studies indicate that the relation between reservoir properties and task performance remains only partially understood~\cite{GKLY_Chaos_20,HKJL_OME_22,HKLJ_NP_23,JL_NCE_24}. Notably, Ref. \onlinecite{PVM_FiP_19} discussed the combined and individual potential of nonlinearities provided by the input layer, reservoir bulk and readout layers. Yet, a systematic investigation of how the choice of nonlinearity impacts reservoir performance across different tasks remains to be carried out and constitutes one objective of this work. Approaching a more interpretable RC scheme, Gauthier et al. \cite{GBGB_NC_21} replaced the complex random network with vectors of the present and past inputs (linear part) as well as their products (nonlinear part) that are linearly combined into a polynomial in the output layer. The resulting model was identified with a \emph{multi-step integration scheme} \cite{I_CUP_09}, using past values of the solution to better approximate the derivative in the interval of integration at high orders of accuracy. Our results indicate that a passive linear setup coupled to a nonlinear readout as provided by a PD can lead to a similar scheme in the presence of transient coupling or time-delayed feedback.

To obtain such an interpretable description, we compare the learned dynamics of the reservoir with the polynomial maps underlying different benchmark tasks derived from two distinct continuous dynamical systems, namely the  \emph{Lorenz63} system \cite{L_JAS_63} and the \emph{Chua oscillator} with cubic nonlinearity~\cite{SAR_Chaos_10}.
Numerical integration of any continuous dynamical system yields a time-discrete approximation in the form of an iterated map derived from the original differential equations. This map can be written in each component as a polynomial of the input. Here, a \emph{polynomial} is defined as
\begin{align}
	\mathbf{P}=\sum_i k_i X_{1}^{\nu_{i,1}}...X_{n}^{\nu_{i,n}},
\end{align}
where $X_1,...,X_n$ are the state variables, $k_i$ are the coefficients and $\nu_{i,n}\in\mathbb{N}$ are non-negative integer exponents. Each summand of $\mathbf{P}$ is composed of so-called \emph{monomials} $kX_1^{\nu_1}...X_n^{\nu_n}$ of a degree defined as the integer sum $\nu_1+...+\nu_n$. The degree of $\mathbf{P}$ thus equals the maximum of the degrees of the contributing monomials \cite{L_Algebra_02}. The polynomial map underlying our tasks was obtained using the symbolic computing python-package SymPy \cite{MSP_PCS_17}. Thus calculating the coefficients of different monomials, allows us to study relative importance of different nonlinear contributions to a task.

Since our readout itself implements a polynomial transformation, the trained reservoir can likewise be represented as a polynomial map. This enables a direct comparison between the learned and target monomials. We identify, which monomials are required by the task and how the nonlinear readout, transient coupling and delayed feedback contribute individually and jointly to the computation of those terms. While in some situations the learned polynomial closely reproduces the task contributions, we also identify qualitatively different computational strategies, where the transient dynamics and delay compensate for missing nonlinear terms, providing new insight into the operation of passive photonic reservoir computers.

This paper is organized as follows: In Section~\ref{sc:model}, we introduce the proposed model, define the corresponding parameters, and describe how a discrete-time input time series is masked and injected into the device. Section~\ref{sc:tasks} presents the considered benchmark tasks together with the underlying dynamical systems. The results are organized into three parts. First, in Section~\ref{sc:separated_nodes}, we analyze the model in the limit of well-separated virtual nodes, i.e., in the absence of inter-node interactions. Next, Section~\ref{sc:transient} incorporates transient dynamics and investigates their individual impact on the reservoir's performance. These findings are then compared with the case of time-delayed feedback in Section~\ref{sc:time-delay}. While the preceding sections primarily focus on short-term prediction measures, the final Section~\ref{sc:reconstruction} examines long-term dynamics and attractor reconstruction capabilities for all previously discussed cases, thereby concluding the study.

\subsection{Reservoir model}
\label{sc:model}
\subsubsection{Minimal optical reservoir}
For our passive linear photonic reservoir we propose a simple ordinary differential equation model, that can be interpreted as a basic model for an injected optical micro-cavity of a round trip $\tau_c\ll 1$, bounded by two opposing distributed Bragg mirrors (DBRs) with reflectivities $r_{1,2}$. A corresponding sketch is shown in Fig.~\ref{fig:1}~(b), where the different components are color coded, to display their roles in comparison to the conceptual RC sketch in Fig. \ref{fig:1}~(a)
Under these simplifying hypotheses a model for a normalized slowly varying field envelope $E(t)$ in the micro-cavity reads
\begin{align}
	\dot{E}&=[-1-\alpha-i\delta]E+h\tilde{Y}_0\label{TGTI_1}\\
	v&=|E|^2\label{TGTI_2},
\end{align}
where $v$ is the output signal as detected by, e.g. a PD, which simultaneously acts as a quadratic nonlinearity. The system is driven by
a monomode injection beam with amplitude $Y_0$ and frequency $\omega_0$. The injection is detuned from the micro-cavity resonance $\omega_c$, resulting in the detuning $\delta=\omega_c-\omega_0$.
While previous works such as Refs.~\onlinecite{VMV_NC_14, VDS_Optica_15} pointed out the performance advantages provided by the case of coherent systems, we restrict our discussion to $\delta=0$, reducing the complex field-envelope $E$ to a real field. Despite a slight increase in performance with non-zero detuning, we did not observe a substantial impact of the phase dynamics on the results, presented in this work.

Note too that $\tilde{Y}_0$ represents the injection beam in the presence of an optical intensity or phase modulation \cite{HSJ_arXiv_26}, though we only consider the first. We assume that the rate of variation is sufficiently small as compared to the micro-cavity mode-spacing so that the monomode approximation is justified.
We define $1+\alpha$ as the total losses, where the coefficient $+1$ represents the amount of light leaving the cavity through the partially reflective mirror $r_1$ while $\alpha$ includes the possibility of additional losses due to residual absorption in the medium composing the cavity. 

The micro-cavity is a so-called \emph{vertical Gires-Tournois interferometer}~\cite{GT-CRA-64,SPV-OL-19,SJG-OL-22,SGJ-PRL-22,KSG-OL-22,SKJGW_Chaos_23}, using $r_2 \approx 1$ for the  bottom mirror resulting in $h=2$ for the so-called the intra-cavity coupling factor  $h=(1-|r_1|)(1+|r_2|)/(1-|r_1\,r_2|)$. Finally, the injection $Y_0$ is modulated according to $\tilde{Y}_0=Y_0(1+\gamma J(t))$ and $J(t)$ is the time-multiplexed input with a randomly generated input mask~\cite{HKJL_OME_22}. We note that real or purely imaginary values of $\gamma$ correspond to intensity and phase modulation \cite{HSJ_arXiv_26}, respectively. However, we only consider purely real values of $\gamma$.
\subsubsection{Extension of the reservoir model}
In order to study the impact of time-delayed feedback, we can extend the model~\eqref{TGTI_1}-\eqref{TGTI_2} to
\begin{align}
	\dot{E}&=[-1-\alpha - i\delta]E+hY,\label{eq:feedback1}\\
	Y&=\eta e^{i\varphi}[E(t-\tau)-Y(t-\tau)]+\sqrt{1-\eta^2}	\tilde{Y}_0, \label{eq:feedback2}\\
	v&=|E|^2.\label{eq:feedback3}
\end{align}
by coupling the micro-cavity to a long external cavity with roundtrip time $\tau$, closed by a feedback mirror with reflectivity $\eta$ and phase $\phi$. Here, $Y$ is the slowly varying field in the external cavity. The total accumulated round-trip phase of the delay thus becomes $\varphi=\phi+\omega_0\tau$ which will be set to zero in this paper. A sketch of the complete model is shown in Fig.~\ref{fig:1}~(c). Note that the vertical nature of the corresponding experimental setup might also allow for spatial or polarization multiplexing, that could even further enhance the promising low-energy consumption of a reservoir computer by enabling parallel task processing as discussed in e.g. Ref.~\onlinecite{GXZ_IEEE_20}.
We stress that, in the absence of any forcing like e.g. time-delayed feedback and modulated optical injection, the system~\eqref{TGTI_1}-\eqref{TGTI_2} will exhibit only transient dynamics before reaching a steady state.
\subsubsection{Masking process}
Figure~\ref{fig:1}~(a) illustrates the masking process of a time-discrete time series $\textbf{u}$, where $u_i\in \textbf{u}$ originally corresponds to the point measured at time $i\cdot \Delta t$ in the time domain domain of the dynamical system. $\Delta t$ is here the sampling step, i.e. the separation between data points sampled for the task. The random input mask $\textbf{M}(t)$ is a piecewise constant function with $N$ plateaus of duration $\theta$ summing up to $T$, which repeats periodically. The mask is multiplied onto each input data point $u_i$, creating the $N$ virtual nodes. The values of the mask levels are drawn from the uniform distribution in the range $[-0.1,0.1]$. The resulting input stream $J(t)$ is then fed into the reservoir. Since an input point is multiplied to all $N$ plateaus of the input mask, each point is injected into the reservoir for a total duration of $N\cdot \theta=T$ over the interval $t\in\left[(i-1) T,\,i T\right[$ with the input weights changing after each interval of length $\theta$. Hence, after the masking process, the distance between the input points has been stretched from $\Delta t$ to $T$, as indicated in the top left panel of Fig. \ref{fig:1}~(a).  

While our model has purely linear nodes, nonlinearity is applied to each node at the readout layer. Hence, nonlinearity in the models and the occurring contributions can be easily tracked and discussed. This point is represented in the sketch of Fig.~\ref{fig:1}~(a), where the nonlinear transformation is denoted with a solid orange line, acting as an additional nonlinear (or activation) layer after the reservoir, hence leaving the nodes themselves truly linear. 
To avoid confusion, the different time scales involved in this work are shortly summarized below:
\begin{itemize}
	\item $dt$ (integration step) is the time step originally used to integrate the dynamical system. 
	\item $\Delta t$ (sampling step) is the separation between sampled data points of the task in the time of the underlying dynamical system. If $dt=0.001$ and $\Delta t=0.01$, every $10$th point of the original timeseries is sampled for the task.
	\item $T$ (input clock cycle) corresponds to the total injection time for each distinct data point in the time domain of the reservoir.
	\item $\theta$ (node separation) is the temporal distance between the $N$ virtual nodes, i.e. different linear combinations of an input point due to the input masking, in the time domain of the reservoir. $N\cdot\theta=T$.
	\item $\tau$ (delay) in the reservoir time domain. Its length relative to $T$ and $\theta$ determines how the virtual nodes and input clock cycles are coupled.
\end{itemize}
\subsubsection{Reservoir training}
Our RC implementation is completed with a \emph{ridge regression}, that creates a balance between minimizing the error and handling the size of the weights. There, the quantity to minimize is
\begin{align}
	\epsilon=\|\textbf{W}_\text{out}\textbf{v}-\textbf{u}\|^2+\beta\|\textbf{W}_\text{out}\|^2,\label{eq:this_expression}
\end{align}
where $\textbf{u}$ is the target data matrix and $\textbf{v}$ the collected outputs. Equation \eqref{eq:this_expression} can be solved by
\begin{align}
	\textbf{W}_\text{out}=\textbf{u}\textbf{v}^T\left(\textbf{v}\textbf{v}^T+\beta\mathbb{I}\right)^{-1},
\end{align}
yielding $\textbf{W}_\text{out}$ depending on the regression parameter $\beta$.

\subsection{Tasks}
\label{sc:tasks}
To analyze the reservoir computing potential of linear photonic systems with a quadratic output~\eqref{TGTI_1}-\eqref{TGTI_2} we construct several tasks requiring different nonlinear transformations employing the Lorenz63 system \cite{L_JAS_63} and the Chua oscillator~\cite{SAR_Chaos_10}, both operating in a chaotic regime.

The equations for the Lorenz63 system are
\begin{align}
	\frac{dx}{dt}&=\sigma_l(y-x),\label{eq:Lx}\\
	\frac{dy}{dt}&=x(\rho_l-z)-y,\label{eq:Ly}\\
	\frac{dz}{dt}&=xy-\beta_l z\label{eq:Lz},
\end{align}
with the standard parameter values $(\sigma_l,\rho_l,\beta_l)=\left(10,28,8/3\right)$. The Chua oscillator \cite{SAR_Chaos_10} can be written as
\begin{align}
	\frac{dx}{dt} &= \alpha_c \left( y - (1 + c)x - a x^3 \right), \\
	\frac{dy}{dt} &= x - y + z, \\
	\frac{dz}{dt} &= -\beta_c y - \gamma_c z,
\end{align}
where we used $(\alpha_c,\beta_c,\gamma_c,a,c)=(45,70,0.4,0.03,-1.2)$ .
To better compare the two dynamical systems, the Chua oscillator parameter set was chosen to have a similar leading Liapunov exponent $\Lambda_L$ to the Lorenz63 system for the given parameters, leading to almost identical \emph{Liapunov times} - the inverse of the leading Liapunov exponent - and comparable \emph{valid prediction times} (VPTs).

The VPT is used to evaluate the reservoir's performance on short to intermediate length predictions. It measures the short-term prediction quality by calculating the point where a predicted trajectory first deviates from the true trajectory past a certain threshold. It is defined as in Ref.~\onlinecite{KPWJL_Chaos_23} as
\begin{align}
	 i_v=\mathrm{max}\left\{i|\delta_i<0.4\right\},\quad \delta_i=\frac{|u_i-\tilde{u}_i|^2}{\braket{|u_i-\braket{\tilde{u}_i}|^2}},
\end{align}
with $u_i$ and $\tilde{u}_i$ the true and predicted data points and $\braket{X_i}$ denotes the average of the quantity $X$ up to the point $i$ while $|\cdot|$ is the Euclidean norm. At the point $i_v$ the error $\delta_i$ first surpasses the threshold of $0.4$. The resulting VPT in Liapunov times is obtained as $t_v=i_v\Delta t\Lambda_L$.
Thus, the VPT gives an estimate in Liapunov times how long the trajectory predicted in closed-loop mode, where the predicted points are used as the next input, remains close to the target trajectory within a normalized error threshold.

To create the reservoir computing tasks, the Lorenz63 and Chua system were numerically integrated using a \emph{Runge-Kutta (4)} method. For the Lorenz63 system the integration time-step is $dt=0.001$ and two tasks, Lorenz-$0.01$ and Lorenz-$0.1$, are created by down-sampling the resulting time series by successive decimations.
For the Chua oscillator, an integration time-step of $dt=0.01$ is employed and the time series is used without any down-sampling. In all three cases, all dynamical variables are normalized and supplied as input to the reservoir and the reservoir is trained to predict one step ahead in each of the input time series. In the prediction phase, we operated the trained reservoirs closed-loop mode.

  Our approach of integrating the Lorenz63 system with a high-order numerical method and a very small time-step and then sampling differently the resulting output, creates two tasks that converge well to the underlying dynamical system but also towards each other. The latter feature is good for comparison while still producing greatly different requirements for any neural network approaching these tasks. Even though the underlying dynamical system contains no monomials beyond quadratic degree, the integration scheme produces monomials of higher degree whose scale separation depends on the integration time-step, or equally on the number of integration steps with $dt=0.001$ in between two data points.

Note that to avoid phase jumps in the injection values, $\gamma$ is chosen such that $\tilde{Y}_0>0$ holds in a vicinity of the target attractor. To that end, we can calculate an upper bound for $\gamma$ by solving $\gamma\min\limits_{t} J(t)>-1$, which leads to
\begin{align}
	\gamma<\frac{10}{\max\limits_{t}\left| \textbf{u}_x(t)\right|+\max\limits_{t}\left| \textbf{u}_y(t)\right|+\max\limits_{t}\left|  \textbf{u}_z(t)\right|},
\end{align}
where the factor $10$ comes from the inverse of the maximum mask value and $\textbf{u}_{x,y,z}$ are the components of the data vector for a three-dimensional task such as those in this work. Note that a longer data set of ten million points was used to get a better approximation for the threshold. Assuming normalized tasks, this results in $\gamma_\text{max}\approx1.225$ for the Lorenz63 system and $\gamma_\text{max}\approx1.745$ for the Chua oscillator and we decided to use $\gamma=1.22$ for both tasks for better comparison.

%
\section{Separated nodes and analytical solution}
\label{sc:separated_nodes}
\begin{figure}[t]
	\includegraphics[width=1\columnwidth]{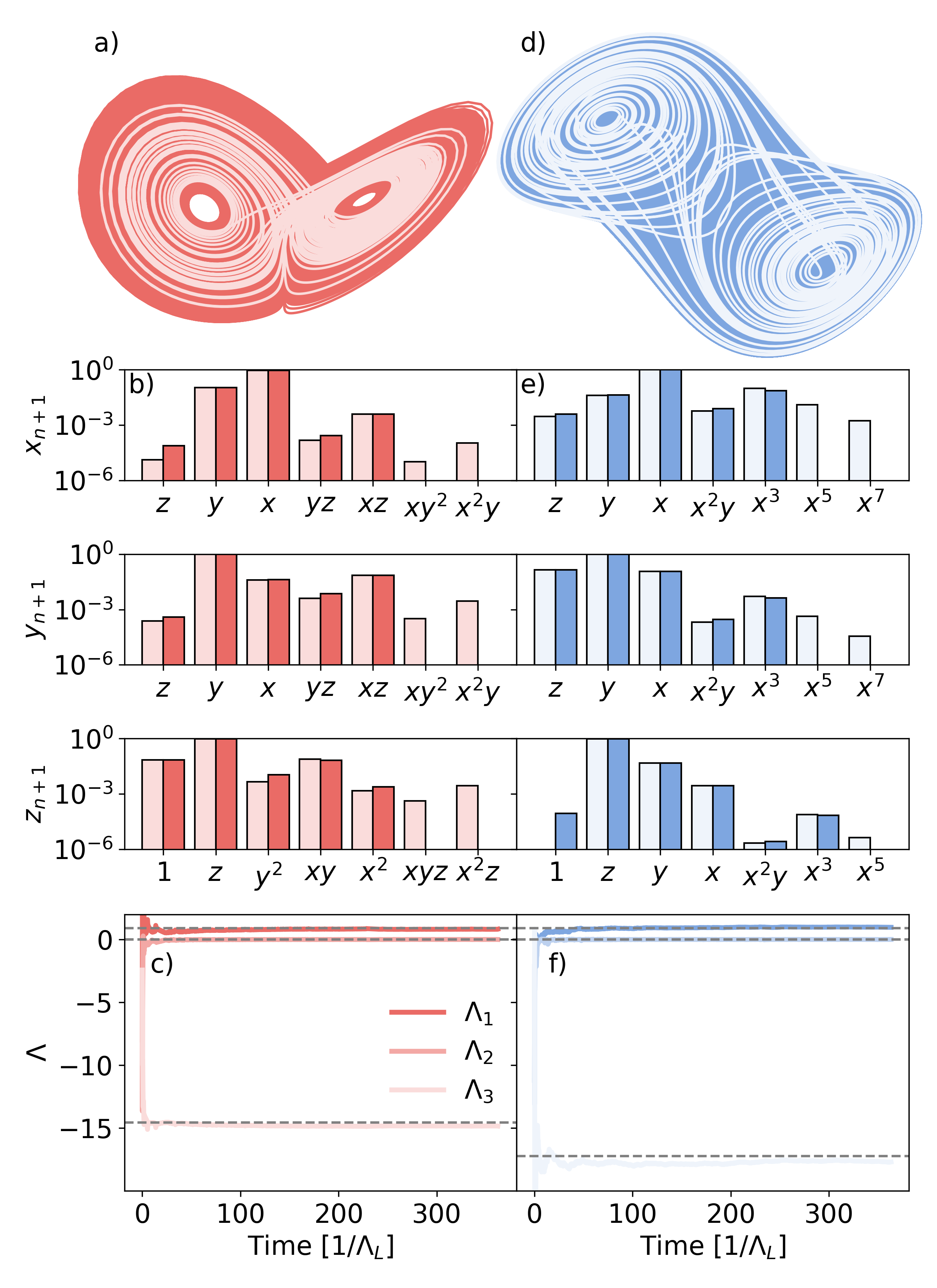}
	\caption{a) Trainings data (bright red) and long-term closed-loop prediction (red) over approximately $2000$ Liapunov times $(1/\Lambda_L)$ for the Lorenz-$0.01$-task with a quadratic nonlinearity in the long input-clock cycle limit without feedback, showing the reproduction of the underlying attractor.
		b) A bar-diagram with the monomials of the task (bright red) and the respective learned map (red). The dominant monomials are well represented, all monomials of higher degree of the task cannot be learned by the quadratic nonlinearity. c) The resulting Liapunov exponents (colored) with the respective values obtained for the underlying dynamical system (gray-dashed).
		(e-f) The same plots for the Chua-task with a cubic nonlinearity.}
	\label{fig:2}
\end{figure}

The simple linear model given by Eqs.~\eqref{TGTI_1}-\eqref{TGTI_2} can be solved analytically. Consider the system is injected with constant amplitude $Y_0$ for a sufficiently long time such that it approaches the steady state at time $t_0=0$. Now the first modulating input value enters the system as $Y_0(1+\gamma j_0)$, where $j_0$ is the first plateau of the input stream $J$. Then, the system \eqref{TGTI_1}-\eqref{TGTI_2} evolves as
\begin{align}
	E(t)&=-\frac{hY_0\gamma}{\lambda}j_{0}e^{-\lambda (t-0)}+\frac{hY_0}{\lambda}(1+\gamma j_0),
\end{align}
with $ \lambda=1+\alpha+i\delta$. At time $t_1$, the injection is changed to $j_1$. Now the system evolves as
\begin{align}
	E(t)&=\frac{hY_0\gamma}{\lambda}\left(j_0-j_1-j_{0}e^{-\lambda t_1}\right)e^{-\lambda (t-t_1)}+\frac{hY_0}{\lambda}(1+\gamma j_1).
\end{align}
This way, we collect sequentially terms of the type $j_{i-1}e^{-\lambda (t_i-t_{i-1})}$. The distance $\theta=t_i-t_{i-1}$ again corresponds to the node separation, given by $T/N$, where $T$ is the input clock cycle and $N$ is the number of virtual nodes, cf. Fig.~\ref{fig:1}~(a). For large values of $\theta$, the remaining terms from the previous injection steps vanish and the system always reaches  the steady state before a new output value is registered. In this far-node-separation limit, where any transient from the time-multiplexing fully decays towards the steady state, our model~\eqref{TGTI_1}-\eqref{TGTI_2} injected with the $i$-th input point $u_i=(x_i,y_i,z_i)\in\textbf{u}$ can be described as a \emph{feedforward neural network} (FNN), reading
\begin{align}
	\textbf{j}=&W_\text{in}u_i,\nonumber\\
	\textbf{v}=&\left|\kappa Y_0(1+\gamma \textbf{j})\right|^2,\label{eq:FNN}\\
	\tilde{u}_{i+1}=&W_\text{out}\textbf{v},\nonumber
\end{align}
with $\kappa=h/\lambda$. In case of $N$ virtual nodes, $\textbf{j}$ contains the $N$ masked input values for the point $u_i$ and $\textbf{v}$ the $N$ resulting output values. Finally, $\tilde{u}_{i+1}$ is our prediction for $u_{i+1}$.
Here, it is obvious that each virtual node describes a polynomial of second degree of the type
\begin{align}v_i^{(n)}=&\left|\kappa Y_0(1+\gamma [w_{nx}x_i+w_{ny}y_i+w_{nz}z_i])\right|^2,
\end{align}
where $n$ refers to the $n$-th virtual node.
The trained output layer $W_\text{out}$ creates a superposition of the $N$ polynomials to predict the next step. Ideally, the learned polynomial converges to the true polynomial or map, connecting the points of data.
This can be seen in Fig. \ref{fig:2}~(a-c) for the model \eqref{TGTI_1}-\eqref{TGTI_2} trained with the Lorenz-$0.01$-task in far node-separation limit at $T=500$ with $N=32$ virtual nodes and thus $\theta > 10$.
In Fig.~\ref{fig:2}~(a), the white trace shows the training set of 6500 data points, while in red we see a closed-loop prediction of $2000$ Liapunov times. The three bar diagrams in b) show the coefficients of the leading seven monomials from the task (bright red), compared to the trained coefficients (red). There, we see that the leading terms fit almost perfectly up to the maximum degree which can be represented by the quadratic nonlinearity, i.e. monomials that are at most quadratic. Indeed, this trained system already produces chaotic dynamics as the Liapunov exponents indicate in Fig. \ref{fig:2}~c), though they slightly differ from the exact Liapunov exponents of the Lorenz system, that are represented by the dashed horizontal lines. 

\begin{figure}[t]
	\includegraphics[width=1\columnwidth]{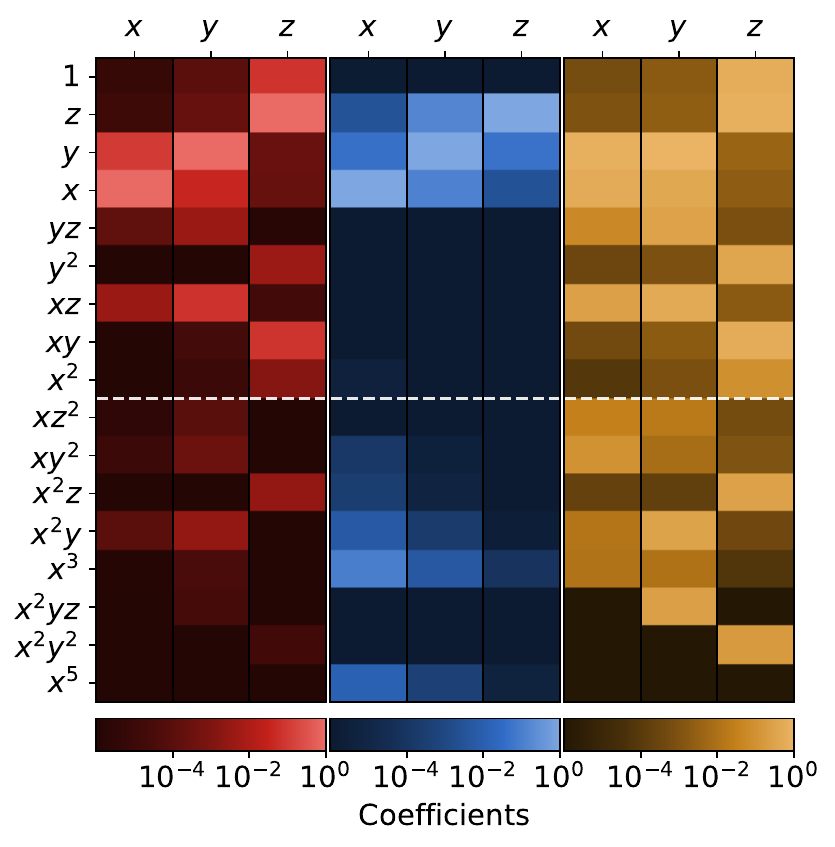}
	\caption{The most influential monomials to the different tasks, split by component. The color indicates the magnitude of the corresponding coefficients. While the Lorenz-$0.01$ (red) and Lorenz-$0.1$-task (yellow) are derived from the same system, the (higher degree) nonlinear contributions are greatly increased in the latter due to the higher integration time-step. For the Chua-task (blue), the most influential nonlinear contributions are at least of third degree due to the cubic nonlinearity. The white dashed line is located at the boundary between quadratic and cubic nonlinear terms. Only for the Lorenz-$0.01$-task the majority of dominant nonlinear contribution is at most of second degree.}
	\label{fig:3}
\end{figure}

We observe a similar result for the Chua oscillator, see Fig.~\ref{fig:2}~(d-f). Note that no decently working example could be found in the far-node separation limit with a quadratic nonlinearity. This can be explained by comparing the dominant monomials from the different tasks, as shown in Fig.~\ref{fig:3}. There, monomial contributions are arranged over the different tasks, split by their respective components. The magnitude of the coefficients are represented by the brightness of the colors. Note that we only show those monomials in the diagram that enter into the leading seven contributions of any component within any of the three tasks. For all the tasks, the most important contributions are the linear ones. 

The major nonlinear contributions to the Lorenz-$0.01$-task (red) are $x\,y$ in the $z$-component and $x\,z$ in the $y$-component. This is evident, considering that these are the respective nonlinear terms in the underlying equation \eqref{eq:Ly},\eqref{eq:Lz}. Similarly, for the Chua-task (blue), the major nonlinear contribution is $x^3$ in the $x$-component, followed by $x^5$ and $x^2\,z$ in the $x$-component too, as well as $x^3$ in the $y$-component. Here, one can see that a squared nonlinearity cannot provide any relevant monomials beyond the linear ones. The relative influence and separation between nonlinear terms within a task is heavily influenced by the integration time-step. This can be seen from the results for the Lorenz-$0.1$-task (yellow) in Fig. \ref{fig:3}, where the large time-step renders almost every single monomial up to a degree of four a relevant contribution.
Below the white-dashed boundary in Fig. \ref{fig:3} we find those terms that are of third degree or beyond and cannot be learned by a quadratic nonlinearity. For the Lorenz-$0.01$-task, the most dominant nonlinear monomials are above the white line, therefore contained in the quadratic nonlinearity, while for the Chua-task, the dominant nonlinear terms are cubic, particularly $x^3$. As a result, of the displayed tasks, only the Lorenz-$0.01$-task can be well represented by our simple model. However, if one was to employ a cubic nonlinearity at the output layer, the Chua-task should be easily accessible as we showed in Fig. \ref{fig:2}~(d-f). 

Similarly, while the Lorenz-$0.1$-task seems to have many very influential contributions beyond a quadratic nonlinearity, the task could be well represented within a quartic nonlinearity. However, this demands a greater number of virtual nodes to yield the necessary degrees of freedom for the required monomial contributions. This demonstrates, how the time-step of a task can reduce the nonlinearity needed as the Lorenz-$0.01$-task can be learned by the simple far node-separation limit model, while the Lorenz-$0.1$-task will not converge with a quadratic nonlinearity, even if time-delay or transient dynamics are included. Indeed, the minimum nonlinear degree a model needs to provide is always equal to the highest nonlinear contribution in the underlying system. Regardless of how small one chooses the integration time-step, the cubic $x^3$ contribution is central to any task based on the Chua-oscillator in order to go beyond trivial linear dynamics.

\section{Influence of transient dynamics}
\label{sc:transient}
\begin{figure}[t]
	\includegraphics[width=1\columnwidth]{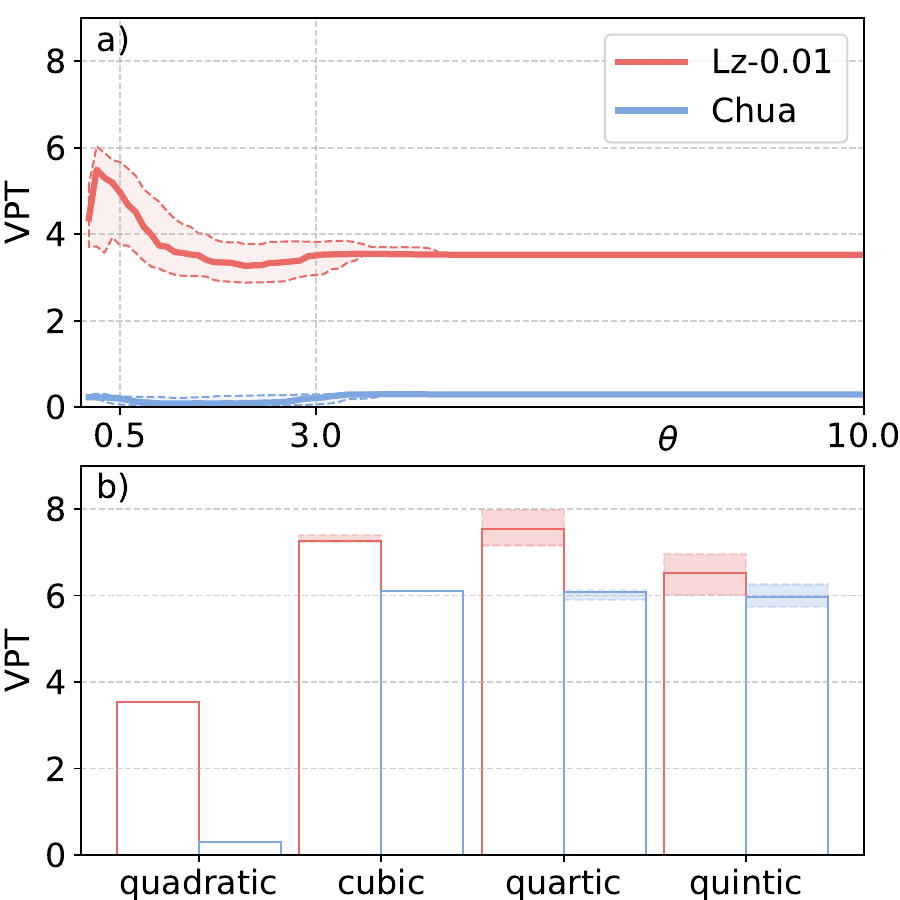}
	\caption{a) The influence of transient dynamics for $N=32$ virtual nodes as valid prediction time (VPT) over varying input clock cycle $T$. The solid lines show the median VPT over $500$ randomly initialized input matrices each evaluated at four different closed-loop starting positions along the attractor, while the dashed lines are positioned at the lower and higher quartiles. For the Lorenz-$0.01$-task at lower $T$ values, a region of higher VPT values can be observed. For higher $T$ values the VPT first drops and ultimately ceases varying. The Lorenz-$0.01$ long-time prediction from Fig. \ref{fig:2}~(a) has been taken from this parameter set with $T=320$. The Chua-task can not be reproduced, though in the presence of interaction at lower $T$ values a small increase in VPT can be observed.
		b) The change of mean VPT in the uncoupled case at $T=320$ and $N=32$. For a cubic nonlinearity, the median VPT of both tasks greatly increase. For a quartic nonlinearity, only the Lorenz-$0.01$ further rises and a variance of results can be observed, indicated by the shaded area. For a quintic nonlinearity, the results of the Lorenz-$0.01$-task even decrease.}
	\label{fig:4}
\end{figure}
The quadratic nonlinearity induced by the PD only allows for a limited amount of monomials for learning the task at hand. Our reservoir is trained to construct the specific linear combination necessary for the polynomial map connecting the points of our trainings data, or rather any map that is sufficiently converging. Increasing the number $N$ of virtual nodes increases the number of degrees of freedom for building the trained map, however, it does not increase the number of monomials available for constructing the map. In this section we shall introduce and discuss the potential uses of including transient dynamics in our model.

If the node separation is not too large compared with the time scales of the reservoir dynamics, an output value can be sampled before the system has fully settled on the steady state for the specific input value. This means that the virtual node output depends on the previous nodes. If instead nonlinear nodes are used, a greater variety of possible reservoir states can be created. With linear nodes however, for the most part, one obtains another linear combination of the input data. The main difference is observed at the end of each input cycle, when the transient dynamics of the last virtual nodes of an input clock cycle influence the first virtual nodes of the next one, creating a memory effect. As our reservoir is linear we can isolate this memory effect, investigating its influence on our trained model. 

In Fig. \ref{fig:4}~(a) the node separation is varied by changing the input clock cycle $T$ for a constant number of $N=32$ virtual nodes. There, we display the VPT over input clock cycle $T$ for the Lorenz-$0.01$-task (red) and Chua-task (blue) for $500$ randomly initialized input matrices and four different closed-loop starting positions along the attractor. The solid line is the median VPT and the dashed lines mark the upper and lower quartiles. For the Lorenz-$0.01$-task at lower $T$ values the highest VPT values occur, indicating a beneficial influence of transient interaction, i.e. memory effects. For higher $T$ values the VPT first drops and ultimately ceases varying, demonstrating the uncoupling of the virtual nodes. The quartiles converging to the median for large $\theta$ means that the different randomly initialized input matrices lead to the same VPT. This indicates that the degrees of freedom for this number of virtual nodes is sufficient to eliminate the influence of the random masks.
The uncoupling of virtual nodes can be demonstrated by reproducing the results from Fig.~\ref{fig:2}~(a) with a FNN of the form of Eqs. \eqref{eq:FNN} for the same input and output matrices, indeed yielding VPTs beyond $25$ Liapunov times.

Finally, while a benefit of transients is evident here, one could wonder, why for the squared nonlinearity, a stronger coupling between virtual nodes should be advantageous to the performance of the Lorenz-$0.01$-task. This is in line with other studies, however these studies usually utilize nonlinear nodes where transient interaction between nodes simply leads to nonlinear dynamics, creating new states and an enhanced feature space. With linear nodes though, the systematic benefit of inter-node interaction can be attributed solely to the memory effects between different input clock cycles around their boundary. As a result, the approximated polynomials will extend from those of the form
\begin{align}
	\left(1+\omega_1 x_i +\omega_2 y_i + \omega_3 z_i\right)	^2 \label{eq:one_step}
\end{align}
towards something like
{\small
	\begin{align}
		\left(1+\omega_1 x_i +\omega_2 y_i + \omega_3 z_i+\omega_4 x_{i-1} +\omega_5 y_{i-1} + \omega_6 z_{i-1}+...\right)	^2, \label{eq:two_step}
\end{align}}
where $\omega_m$ are the full coefficients in the final superposition of the virtual nodes, depending non-trivially on the weights of the input matrix. 
While it is important to point out that doing so does not increase the monomial degree (a term like $x_i^2y_i$ still could not be found here), it does yield an advantage; while a method such as Eq.~\eqref{eq:one_step} could fully represent, for the Lorenz63 system, an explicit Euler step
\begin{align}
	u_{i+1}=u_i+\Delta t f(u_i),
\end{align}
which has a first global order in time-step $dt$, the latter Eq.~\eqref{eq:two_step} allows us to fully reproduce all polynomials created by an explicit multi-step method such as a two-step Adams-Bashforth (2) scheme
\begin{align}
	u_{i+1}=u_i+\frac{3}{2}\Delta t f(u_i)-\frac{1}{2}\Delta t f(u_{i-1}).
\end{align}
Going from an Euler scheme towards a Runge-Kutta method of order $n$, the algorithm increases accuracy at the cost of the need to compute increasingly higher degree monomials. Meanwhile, a multi-step method of order $n$ such as the $n$-th order Adams-Bashforth can be constructed from the present as well as the previous $n-1$ input points~\cite{I_CUP_09}, without the need of any monomials beyond the highest degree monomial of the underlying dynamical system, e.g. quadratic in case of the Lorenz63 system. A similar scheme has been used for next generation reservoir computing \cite{GBGB_NC_21}, where the input vector from current and previous steps has been included to construct the feature vector and multi-step integrators were used to explain the good performance. Note that we can verify that we obtain a multi-step method from coupled virtual nodes, by confirming that the previous step enters the prediction. Yet, due to the complex coupling, the exact impact is hard to quantify such that we cannot identify the specific multi-step method or the numerical order of accuracy.

There are also limitations for the learning of multi-step methods through transient coupling. First of all, they mainly include the previous input clock cycle and often only the first few virtual nodes of an input clock cycle are coupled to the previous cycle. Additionally, the combinations of previous and present inputs are limited due to the fixed time-order of nodes as we cannot just combine the $n$-th node of the previous input clock cycle with the $m$-th node from the present one, without all nodes in-between entering in a fixed order. These aspects can be greatly enhanced by the inclusion of time-delayed feedback, as we will see in the next section.

Note that in Fig. \ref{fig:4}~(a) the Chua-task cannot be learned due to the limited squared nonlinearity, though in the presence of transient interaction at lower $T$ values a small increase in VPT can be observed. This can be attributed to the corresponding reservoirs learning a periodic orbit with a similar transient flow to the underlying attractor.

To characterize the influence of the nonlinear order, we implemented different nonlinearities $v=|E|^p$ , beyond the quadratic one used in Eq.~\eqref{TGTI_2}, with indices $p=3,4$ and $5$. The results are shown in Fig.~\ref{fig:4}~(b) for the Lorenz-$0.01$-task (red) and the Chua-task (blue) with $N=32$ at $T=320$ in the far-node separation limit. For the Lorenz-$0.01$-task, the optimum performance seems to be achieved for $p=4$, which can be understood considering that most relevant terms are at most of quartic polynomial degree. Interestingly, we observe a decline in performance for $p=5$. 

Though not shown here, this effect shrinks and ultimately vanishes for a larger number of virtual nodes and might be explained with too small number of virtual nodes to handle all the possible monomials which can occur for $p=5$ (particularly because some degrees of freedom need to be attributed to canceling out the new fifth degree terms, that are almost all irrelevant to the Lorenz-$0.01$-task). For a quartic nonlinearity we get $35$ monomials, which is slightly above the output dimension, and hence can not be fully sampled. This can explain the variance between upper and lower quartiles starting at quartic degree, as the results now depend on the input matrix. With $56$ possible monomials at $p=5$ the number of degrees of freedom in the output is greatly below the requirement, leading not only to a larger variety of resulting VPT values, but also to a decreased median value as the many unnecessary monomials often interfere with the performance of the trained network.
For the Chua-task, an increase in performance can be seen at third degree. Here, a similar effect of raising variance can be observed at $p=4$ and $p=5$, however on a greatly reduced scale, possibly due to the Chua-task effectively only using a very small number of monomials.
\begin{figure}[b]
	\includegraphics[width=1\columnwidth]{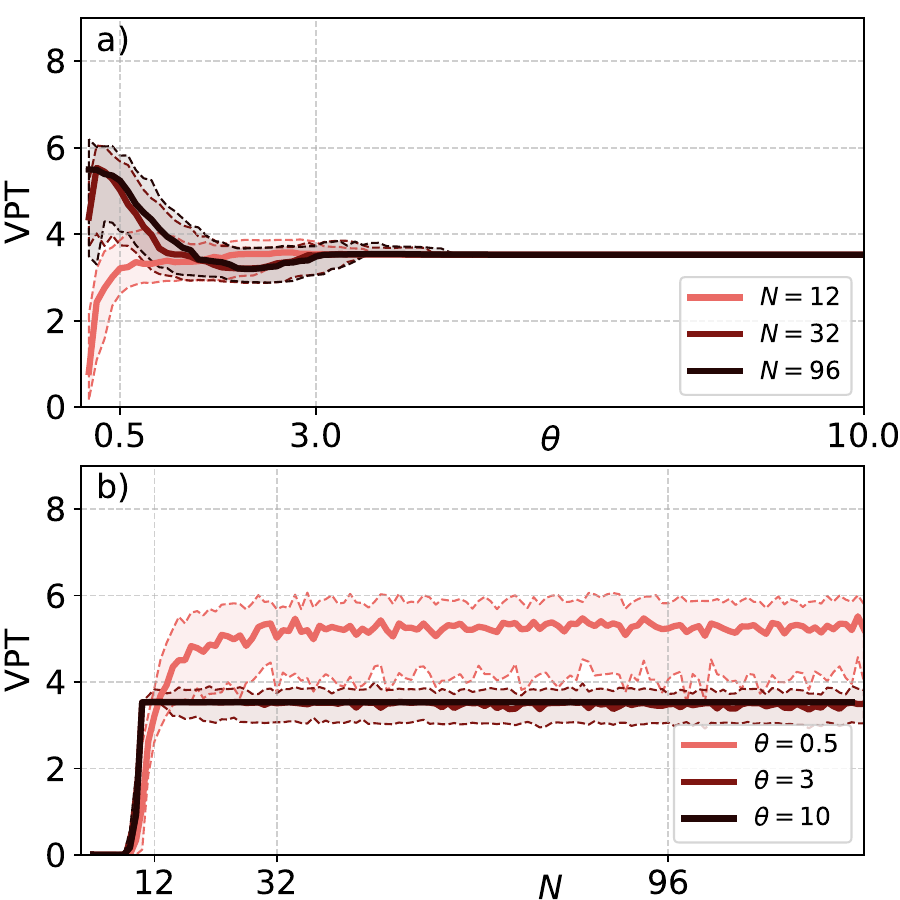}
	\caption{a) VPT over virtual node separation $\theta$ in the absence of delay for different virtual node numbers $N$. The solid lines correspond to the median over $100$ randomly initialized input matrices and four different closed-loop starting positions along the attractor. In the uncoupled regime, neither $N=32$, nor $N=96$ virtual nodes poses any advantage over $N=12$. In the presence of transient coupling, $N=12$ virtual nodes fall back compared to the higher numbers. This becomes clearer with b), showing VPT over the number of virtual nodes $N$ for different virtual nodes separation values $\theta$. In the uncoupled case for $\theta=10$ a step can be observed around $N=10$ from no convergence to maximum performance, which is exactly the number of monomial terms that can be produced by the square nonlinearity. Including transient coupling at lower node separation gradually increases the median VPT, while the arrival at the maximum VPT plateau is also shifted to higher $N$ values.}
	\label{fig:5}
\end{figure}

The connection between the number of possible monomials, the number of virtual nodes and the node separation is further illustrated in Fig.~\ref{fig:5}. Figure~\ref{fig:5}~(a) shows VPT as a function of $\theta$ for different numbers of virtual nodes $N$ where solid line marks the median from $100$ randomly initialized input matrices and four different closed-loop starting positions along the attractor. Here, we can see that in the far node-separation limit all lines converge, as all networks provide a sufficient number of degrees for freedom for the uncoupled case. In Fig. \ref{fig:5}~(b) for $\theta=10$ (black), the uncoupled case, we can see a sharp jump in VPT from zero to maximum around $N=10$, which is exactly the number of monomials offered by the squared nonlinearity. In the presence of strong transient coupling at low values of $\theta$ in Fig. \ref{fig:5}~(a) we can see a systematic benefit of $N=32$ over $N=12$, but no meaningful benefit of $N=96$ over $N=32$, an impression which is strengthened by Fig.~\ref{fig:5}~(b) where the lines for coupled nodes $\theta=0.5$ (bright red) and $\theta=3$ (red) arrive at their maximum plateau at higher values of $N$ than the uncoupled case. This indicates that a few additional degrees of freedom are required to use the benefit provided by transient coupling. Note that the free combination of the present and previous input clock cycle would lead to $28$ monomials and the arrival at a maximum VPT far below this $N$ value indicates the limited combinations yielded from a pure transient coupled case.
\section{Influence of time-delay}
\label{sc:time-delay}
\begin{figure}[b]
	\includegraphics[width=1\columnwidth]{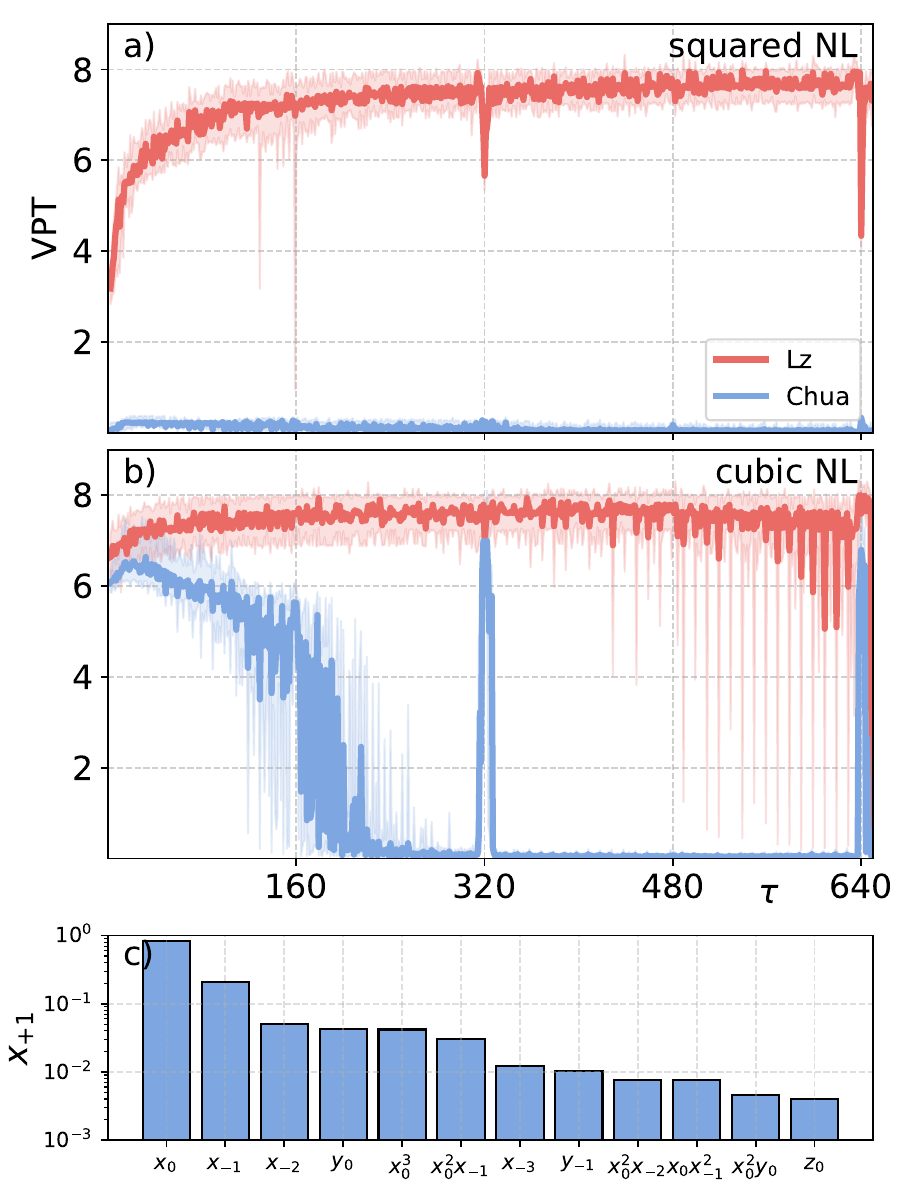}
	\caption{a) Influence of a time-delayed feedback with a feedback strength of $\eta=0.6$ and the qudratic nonlinearity. VPT as a function of the time-delay $\tau$ for $T=320$ is depicted. The solid line shows the median over $100$ randomly initialized input matrices and four different closed-loop starting positions along the attractor. One can observe an increase in VPT with rising $\tau$ beyond the previous results. A notable drop appears around $\tau=T=320$, which is the resonant case, as well as at the second resonance and at $\tau=T/2$. The Chua-task does not profit from the time-delayed feedback. b) Introducing an cubic nonlinearity. Again, the VPT rises for the Lorenz-$0.01$-task. Here, the Chua-task starts at its VPT maximum at $\tau=0$ and gradually falls towards zero. However, the absolute maximum arises around the resonances. c) $x$-component of the map obtained in resonance $T=\tau$ for the Chua-task with a cubic nonlinearity in the limit of far-separated virtual nodes.}
	\label{fig:6}
\end{figure}
\begin{figure}[b]
	\includegraphics[width=1\columnwidth]{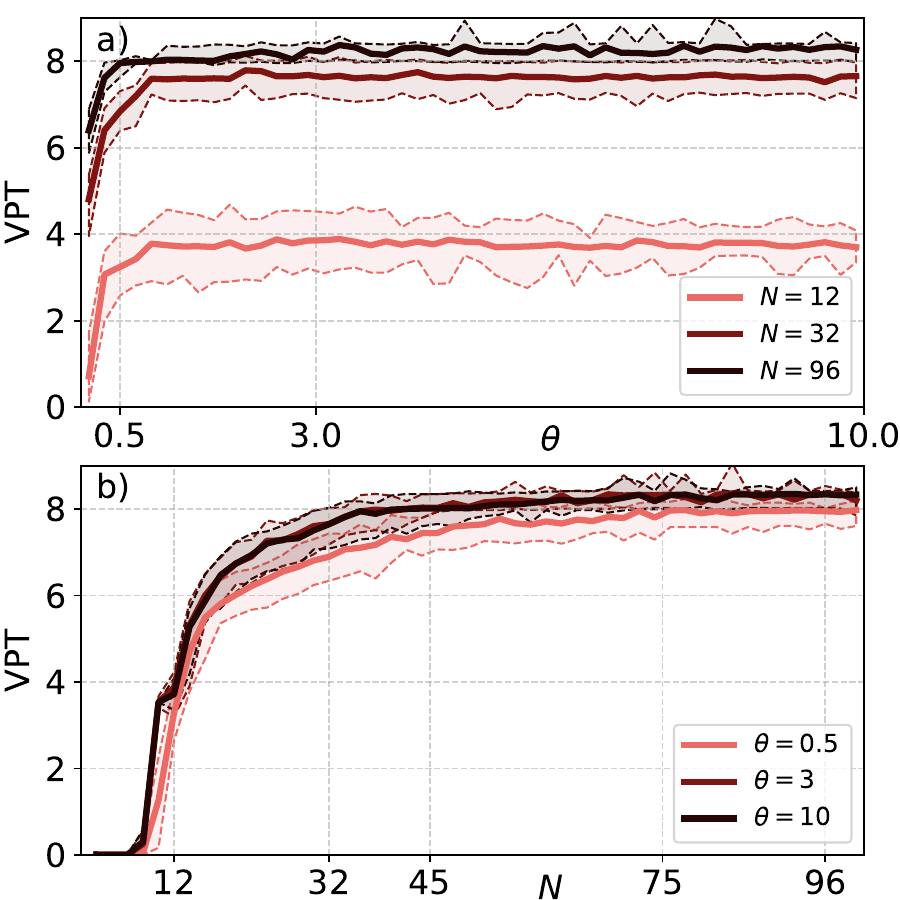}
	\caption{Impact of time-delayed feedback on the connection between the occurring number of monomials, the number of virtual nodes and the node separation (cf. Fig. \ref{fig:5}).
		a) VPT over node separation $\theta$. Now, there is almost no effect of transient coupling as the delay seems to replace it. The representation over the node number $N$ shows a first jump of performance at around $N=10$ and continuously rising VPT values as $N$ is further increased.}
	\label{fig:7}
\end{figure}

In this section, time-delayed feedback is introduced by coupling the previously considered micro-cavity with roundtrip time $\tau_c$ to a long external cavity with roundtrip time $\tau\gg \tau_c$. A sketch can be seen in Fig. \ref{fig:1}~(c) while the respective equations are Eqs. \eqref{eq:feedback1}-\eqref{eq:feedback2}. A system can be time-delayed and linear at the same time. Time-delay can act similarly to transient dynamics - as long as the delay is shorter than the node separation and the feedback strength is relatively large, both effects induce a gradually decaying coupling between neighboring nodes. However, as soon as the time-delay is larger than the node separation, it can skip one or several nodes and if it is chosen larger than the input clock cycle $T$, the connection is exclusively between different data points, acting only as memory. There is a special resonant case $T=\tau$, where each virtual node at the present input clock cycle is coupled to and receives feedback from itself from previous cycles~\cite{KYL_JoP_21}. As we will see, this case tends to show different or even opposing behavior to the general inclusion of time-delayed feedback. 

With the results from previous sections we can already have reasonable expectations on the systems behavior in the presence of time-delay. In particular in the context of multi-step numerical integration methods delay has an advantage over transient dynamics. If the delay time is chosen in an order of magnitude similar to the input clock cycle $T$, it will create a fading memory of the previous $n$ input points that can be used for higher order multi-step methods. This is in contrast to models produced by transient dynamics, where usually only the current and the previous input point are included. As a consequence the learned model in the time-delayed  case can converge closer to the task produced by a fourth order Runge-Kutta algorithm. This is due to the algebraic nature of Eq.~\eqref{eq:feedback2}, including the $n$ previous roundtrips with a strength $\eta^n$. In the limit of weak time-delayed feedback, the memory introduced by the time-delay is rapidly fading. Here, we consider strong time-delayed feedback with values of $\eta\geq 0.5$, where the memory reaches over multiple roundtrips. Also, the delay can be chosen to freely combine any virtual node from the current input cycle with any node from the previous one, leading to a greater variety of reservoir states and the freedom necessary to construct those higher order multi-step methods, given a sufficient number of virtual nodes is provided.

Figure \ref{fig:6}~(a) shows VPT as a function of the delay time $\tau$ with a squared nonlinearity for a feedback strength of $\eta=0.6$, $N=32$ virtual nodes, $T=320$ where the line is the median VPT from $100$ randomly initialized input matrices. A delay between $T$ and $2T$ seems to produce particularly good results for the Lorenz-$0.01$-task (red). An exception occurs around the resonances at $\tau=T$, $\tau=2T$ and $\tau=T/2$, where a drop in VPT is observed, which we attribute to the limited amount of combinations in a situation where every virtual node from the present input cycle is only coupled to itself from the previous cycles leading to reduced freedom for the construction of a polynomial of type Eq.~\eqref{eq:two_step}. Meanwhile, the Chua-task (blue) does not benefit significantly from time-delayed feedback as the quadratic nonlinearity is still below the minimum degree necessary for this task. Yet we observe a slight increase in the median VPT for small $\tau$ values and around the resonance. Studying the corresponding learned models, we often find limit cycles such that the closed-loop trajectory evolves close to the ground-truth for a short time. 

In Fig.~\ref{fig:6}~(b) we can see the same figure for a cubic nonlinearity. Here, the benefit of time-delayed feedback for the Lorenz-$0.01$-task is smaller, due to the performance already being very good without time-delayed feedback. For the Chua-task we observe that the resonant case apparently conserves or even increases performance, while other $\tau$ values lead to reduced VPT values. Thus, the same previously limiting property of virtual nodes only coupling to themselves seems to act as a stabilizing influence. This special case of resonant time-delayed feedback might even be useful for robustness, e.g. in the presence of noisy data. Note that assuming far-separated nodes, our extended model with time-delayed feedback can be once again approximated with an analytical network at the resonance $T=\tau$. As each virtual node is only coupled to itself from previous steps, the memory takes a diagonal form and the resulting model can be written as a map \cite{J_MLST_24}. This can be used to investigate the effect of resonant time-delayed feedback on the learned model for the Chua-task with cubic nonlinearity, yielding
\begin{align}
	\textbf{j}=&W_\text{in}u_i,\\
	\textbf{Y}_i=& \eta(\kappa-1)\textbf{Y}_{i-1} + \sqrt{1-\eta^2}Y_0(1+\gamma \textbf{j}),\label{eq:recursion}\\
	\textbf{E}_i = &\kappa \textbf{Y}_i,\\
	\tilde{u}_{i+1}=&W_\text{out}|\textbf{E}_i|^3.
\end{align}
If initialized equivalently with the same warm-up stage, this reproduces the closed-loop predictions of the corresponding trained reservoir for more then $10$ Liapunov times, leading to identical VPTs. Applying the same monomial decomposition as in Section \ref{sc:transient}, we recover the leading contributions shown in Fig. \ref{fig:6}~c), where $x_{-i}$ denotes inputs from $i$ time steps in the past. 
The dominant terms coincide with those of the corresponding reservoir devoid of time-delayed feedback displayed in Fig. \ref{fig:2}~e), which are $x_0$, $y_0$, $z_0$, $x_0^3$, and $x_0^2y_0$, indicating that the underlying functional structure is preserved.
The additional contributions introduced by the resonant delay take the form of temporally shifted "echo" terms such as $x_{-1}$ and mixed products like $x_0^2x_{-1}$. In agreement with Eq. \eqref{eq:recursion} the coefficients of $x_{-n}$ are smaller than $x_0$ by a factor $(\eta(\kappa-1))^n$. Rather than introducing new dominant nonlinearities, these terms reflect a multi-step replication of existing monomials, mainly the cubic $x^3$ term which is the dominant nonlinearity in the Chua oscillator. This suggests that resonant time-delayed feedback does not fundamentally alter the learned model, but instead reinforces its original functional structure while improving stability.

Finally, we will pick up once again the connection between the possible number of monomials, the number of virtual nodes and the node separation, comparing Fig.~\ref{fig:5} in the presence of transient coupling to Fig.~\ref{fig:7}, now including time-delayed feedback. Figure~\ref{fig:7}~(a) demonstrates how the effect of transient and thus virtual node-separation is effectively eliminated by time-delayed feedback. Simultaneously, contrary to Fig.~\ref{fig:5} there is now a benefit of increasing the virtual node number $N$ beyond $32$, illustrating the larger number of utilized degrees of freedom in the presence of time-delayed feedback compared to the case of transient coupling only. The dependence of VPT on the virtual node number $N$ as well as the reduced dependence on the virtual node separation is further visible in Fig. \ref{fig:7}~(b). There, the median VPT starts with a first performance increase at around $N=10$ of similar magnitude as in Fig. \ref{fig:5}~(b). However, now it continues rising until beyond virtual node numbers of $N>40$. Additionally, a larger node separation seems to be beneficial in the presence of time-delayed feedback in some cases. Finally, we note that not only the VPT trend in dependence of $N$ is changed in the presence of feedback, but the absolute performance is also improved.
\section{Comparing attractor reconstruction capabilities}
\label{sc:reconstruction}
\begin{figure}[b]
	\includegraphics[width=1\columnwidth]{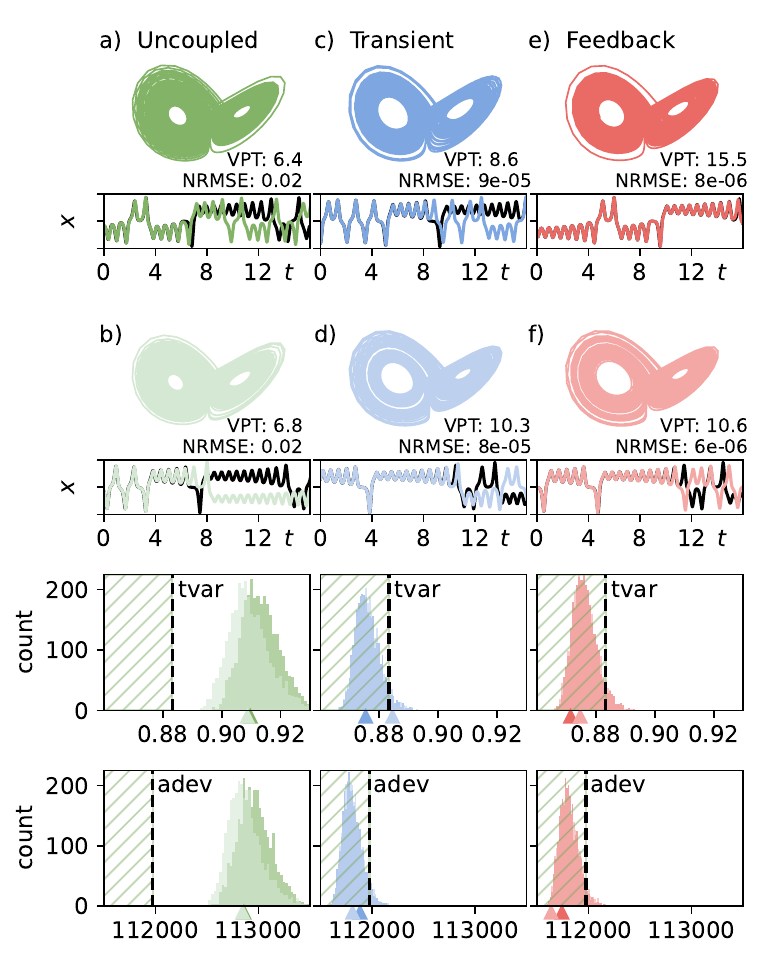}
	\caption{Performance evaluation for six different trained reservoirs with a squared nonlinearity at the output layer, using multiple measures. All panels show a closed-loop prediction observation of $10^4$ steps, below the respective first $15$ Liapunov times of the $x$-component (colorized), superposed with the true trajectory (black) to illustrate the VPT. VPT and NRMSE values are given too. Below, we show the respective test values for $5000$ observables of the same type as a histogram. If the majority of the histogram is below the threshold (black-dashed), the test is accepted. The triangular markers show the test values of the exemplary observations above. (a-b) Trained reservoirs with $32/10$ uncoupled nodes due to high node separation of $\theta=10$. (c-d) Trained reservoirs with nodes coupled through transient dynamics due to a lower node separation of $\theta=0.25$. (e-f) Trained reservoirs with $N=50$ and a node separation of $\theta=10$, only coupled due to time-delayed feedback with feedback strength $\eta=0.6$.}
	\label{fig:8}
\end{figure}
Note that VPT values can vary greatly depending on the random seed or the initial position along the attractor where the closed-loop simulation has been started. To account for this, we have been using a median VPT of a number of trained reservoirs, ranging from the low to mid hundreds, also including multiple different initial positions along the attractor. Yet, a high VPT does not capture the models long-time behavior, i.e. the capability of reproducing the target attractor.
The learned attractor could be deformed with respect to the target or the closed-loop trajectory might settle on some periodic orbit or fixed point after a transient of arbitrary length. Hence, in the final section, we shall introduce two measures capturing the long-term behavior of the system. To that end Ref.~\onlinecite{FLWHJ_arXiv_25} introduced multiple measures, e.g., \emph{total variation} (TVar) and \emph{attractor deviation} (ADev). The TVar is an empirical approximation of the total variation distance between the reference probability density function $\pi(u)$ produced by the Lorenz63 system and the probability density distribution of data produced by the trained reservoir $q(u)$. Here, $u$ in case of the Lorenz63 system is a point in $\mathbb{R}^3$ and the probability density $\pi(u)$ captures how frequent $u$ is visited by trajectories as the Lorenz system is solved with Runge-Kutta (4) scheme. The TVar equation
\begin{align*}
	d_\text{TVar}(\pi,q)=\frac{1}{2}\int_\mathbb{R^3}\|q(u)-\pi(u)\|du
\end{align*}
quantifies the extent to which the two probability density functions and thus the produced attractors coincide. The ADev measures the deviation of the attractor shape between the two systems. This is achieved by comparing the support of the two system, i.e. the area where both probability density functions are non-zero. In Ref.~\onlinecite{FLWHJ_arXiv_25} the measures are used as statistical tests where a time series with a sampling $\Delta t=0.02$ and $5000$ points is considered as an observation. For each test the authors calculated a threshold so that $95\%$ of observations drawn from the reference system would yield test values below the threshold. 

In Fig.~\ref{fig:8} we apply the tests to six different trained reservoirs (a-f). Panel a) shows the results obtained with a trained reservoir with uncoupled nodes. On the top, we can see a three-dimensional attractor reproduction (i.e. observation) created by a closed-loop simulation over $10^4$ steps with $\Delta t=0.01$. Below, the first $15$ Liapunov times of the $x$ component of the same closed-loop prediction are shown to illustrate the short-term behavior (green) in comparison with the ground-truth (black). The respective VPT and NRMSE values are given too. In panel b), we can see another example with uncoupled nodes, colorized in a brighter shade of green. Further down the column, the respective TVar and ADev distributions for $5000$ observations obtained from different initial points are presented for both exemplary reservoirs with uncoupled nodes, colorized accordingly. The green arrows at the bottom of the histograms mark the respective test values produced by the specific observations shown in (a-b). Even though attractor reproductions of the uncoupled reservoirs look similar to the Lorenz attractor and produced decent VPT values, the tests clearly reject them as observations of a Lorenz63 system. 

By further comparison, we noticed that observations obtained from the uncoupled reservoirs particularly struggle to reproduce the outmost regions of the Lorenz63 system, often staying longer on the inner circles. Even though they are able to produce chaotic behavior with similar Liapunov exponents to the original Lorenz63 system, c.f. Fig. \ref{fig:2}~(c), the resulting attractors are different. Within those badly reproduced outer regions of the attractor, the higher degree nonlinear monomials have a greater contribution, so this is a plausible result. Additionally, we used $N=32/10$ in panels (a/b) to demonstrate that $N=10$ virtual nodes are indeed sufficient to get the full potential of the uncoupled case (c.f. Fig. \ref{fig:5}~b)). In panels (b-c) and (e-f) we can see two observation for the transient (with $N=32$) and time-delayed feedback (with $N=50$) case, respectively. Both cases have around $95\%$ acceptance rate of their observations, hence both can be said to reproduce the Lorenz attractor. However, the time-delayed feedback observations have lower NRMSE values and thus a tendency to produce higher VPT values, even sometimes beyond VPT=$15$, as in panel (e). 
Note that the transient case corresponds to a two-step integration method, while the time-delayed case has access multiple previous steps. We find that both algorithms can be of sufficient numerical accuracy to reproduce the Lorenz attractor and pass the long-term behavior TVar and ADev tests. However, the time-delayed case outperforms the transient case in the short-term performance (i.e., VPT and NRMSE) due to closer converging to the original fourth order Runge-Kutta method generated map of the task.

\section*{Conclusion}

We numerically and analytically analyzed the computational potential and shortcomings of a passive linear optical system, using a PD and thus a quadratic nonlinearity at the optical-to-electrical interface as the only source of nonlinearity. By explicitly calculating the contributing monomials for different tasks, we demonstrated the task requirements and necessary nonlinearity to learn various tasks. The nonlinear monomials appearing in the underlying dynamical system are always dominant in the task and act as a lower limit for the nonlinearity required for the reservoir computer. Secondary nonlinear contributions, induced by the integration algorithm, are reduced in their influence by choosing a smaller time-step. The maximum number of virtual nodes necessary is shown to equal the number of monomials provided by the nonlinearity.
For a dynamical system such as Lorenz63, where the present state fully determines the future, no delay and transient interactions are strictly necessary. 

Hence, a passive and linear reservoir with a PD acting as quadratic output layer, can perform well on a task derived from the Lorenz63 system, given the time-step is sufficiently small. Here, time-delayed feedback can greatly increase the performance, compensating for the missing higher degree monomial terms due to the implicit implementation of higher order multi-step methods by the reservoir, without the need for nonlinearities of higher degree. To fully utilize this benefit, however, a sufficient number of virtual nodes is necessary. We observed that transient coupling can partly produce the same effect. This is particularly interesting for RC realizations without an intrinsic memory mechanism. 

On the other hand, our results indicate that combining transient coupling and feedback effects might even worsen performance in a realization with a limited number of virtual nodes. In particular for physical RC realizations we thus suggest to carefully consider the intrinsic memory and transient properties of the studied setup. While transient coupling and time-delayed feedback both perform well at long-term attractor reconstruction our findings suggest that time-delayed feedback performs better at short-term predictions, i.e. yields higher VPT and lower NRMSE values. 
Meanwhile, a task derived from a system such as the Chua oscillator with cubic nonlinearity, cannot be represented by the system, unless electronic nonlinearities of higher degree are supplied, even in the presence of transient or time-delayed feedback coupling. 

Note that coherent systems with $\delta\neq 0$ might yield richer nonlinear dynamics due to interference imposed by the cosine-term, which emerges from a phase difference between present and previous contributions in the transient and time-delayed case. However, despite a slight improvement in performance, we did not find a different scenario in this situation, given the system is still operated at steady state. In particular, the richer nonlinearity introduced by the  cosine-term does not remove the observed lower limit for the required nonlinearity. Yet it's impact could be interesting for further studies. 

Another promising topic is task multiplexing, i.e., the simultaneous execution of multiple tasks within a single trained reservoir, analogous to the ability of biological neural systems to perform several tasks concurrently. While predicting two tasks with a reservoir of twice the size may not offer an advantage over using two independent reservoirs, computational savings could be achieved if the tasks share a subset of virtual nodes. In this context, the monomial-based analysis developed in this work may provide a useful framework for identifying common monomial contributions between different tasks and, consequently, estimating a lower limit for the number of virtual nodes required for their simultaneous realization.

The analysis presented in this paper is demonstrated on equations describing an optical reservoir with a photodector. However, our insights can also be applied to other systems. The main insights that are gained pertain to the minimal required polynomial order of the reservoir 
 and how transient dynamics and time-delayed feedback can increase the number of monomials provided by the reservoir. That the reservoir must contain the minimal polynomial order required by the task, and that insufficient polynomial degree can not be compensated for by memory, can be used to either exclude physical reservoirs that can not perform the required transformations or to specifically design the reservoir to include the necessary polynomial degree. For optical systems such system design could ,besides electric nonlinearities, include various nonlinear optical elements or even operating the photodector such that higher order nonlinearity can be included via saturation effects. This is however contingent upon sufficient knowledge of the task requirements. The knowledge of how transient dynamics can influence monomial contributions, can in general be used to optimize the size of the reservoir output layer.

\section*{Author Declarations}
The authors have no conflicts to disclose.
\section*{Acknowledgments}
L.J. acknowledges funding from the Carl Zeiss Foundation.
J.J. acknowledges the financial support of the project
KEFIR/AEI/10.13039/501100011033/ FEDER, UE. J.J., E.R.K and
S.V.G. acknowledge the financial support of the project
KOGIT, Agence Nationale de la Recherche (ANR-22-CE92-
0009), Deutsche Forschungsgemeinschaft (DFG), Germany
via Grant No. 505936983 and Nr. 524947050
\section*{Data Availability}
The data that support the findings of this study are available from the corresponding author upon reasonable request.
%



\end{document}